\documentstyle[twocolumn,epsf,aps]{revtex}
\draft
\begin{document}
\title{Branching annihilating random walk on random regular graphs}
\author{Gy\"orgy Szab\'o}

\address
{Research Institute for Technical Physics and Materials Science \\
P.O.B. 49, H-1525 Budapest, Hungary}

\address{\em \today}

\address{
\centering{
\medskip \em
\begin{minipage}{15.4cm}
{}\qquad The branching annihilating random walk is studied on a
random graph whose sites have uniform number of neighbors ($z$).
The Monte Carlo simulations in agreement with the generalized mean-field
analysis indicate that the concentration decreases linearly with the
branching rate for $z \ge 4$ while the coefficient of the linear term
becomes zero if $z=3$. These features are described by a modified
mean-field theory taking explicitly into consideration the probability of
mutual annihilation of the parent and its offspring particles using
the returning features of a single walker on the same graph.
\end{minipage}
}}
\maketitle

\narrowtext

The branching annihilating random walk (BARW) \cite{BG} is considered as
one of the simplest models of the extinction processes exhibiting critical
behavior in different physical, chemical, biological and economical
systems \cite{MD}.
In these phenomena the walkers can represent domain walls, vortices,
defects, atoms, active sites, biological species or their colonies,
strategies, etc. This is the reason why the BARWs have been extensively
studied in the last years (for references see the papers by Cardy and
T\"auber \cite{CT96,CT98}).

In general, the walkers (henceforth particles) jump randomly on one of the
neighboring sites of a lattice and each one can create additional particles
with a branching probability $P$.
Furthermore, two particles annihilate each other if they try to share a
site as consequence of the mentioned jump or branching events. When
varying the branching probability a phase transition can be observed
in the average concentration of the particles. Namely, the particles survive
if the branching rate exceeds a critical value $P_c$, otherwise the system
tends toward the absorbing state (no particles) which is independent of
time. The transition from the active to the absorbing state belongs to the
directed percolation (DP) universality class as well as the
extinction processes in most of the one-component system \cite{dpconj}.

On the lattices the BARWs are well investigated by different techniques
\cite{CT96,CT98,J93,IT,TT}. The Monte Carlo (MC) simulations are
extended to Sierpinski gasket by Takayasu and Tretyakov \cite{TT}. In the
present paper we will study the BARW on random regular graphs characterized
by a uniform number of joints. The joints of these graphs
define the possible paths for the particles in the structureless systems
where the spatial position of the sites is no longer relevant.

It is emphasized that the investigation of some physical
phenomena on graphs provides a more general understanding. For example,
the extension of the Mermin-Wagner theorem to graphs shows that the
recurrence criterion for the absence of continuous symmetry breaking
remains valid in the graphs too \cite{Cassi92,MW,BCV}. In other words,
the existence of the spontaneous magnetization on a graph is related to
the probability of returning to the starting point for a single random
walker on the same graph. The recurrence of a random walk plays also a crucial
role in the BARW because a particle and its offspring will be annihilated
when they meet. The variation of the distance between them can be mapped to a
single walker problem on the same graph. As a result, if the motion
of the parent and its offspring is not affected by other particles then
the probability of their mutual annihilation equals to those of returning
to the starting site for a simple random walk. 

Our investigation will be concentrated on the graphs consisting of $N$
sites and each site has $z$ joints toward different, randomly chosen
sites (henceforth neighbors) excluding itself.
At a given time each site can be occupied by a single particle or empty.
The time evolution is governed by repeating the following elementary
processes. A randomly chosen particle creates an additional particle on
one of the neighboring site with a probability $P$ or jumps to this site
(with a probability $1-P$). In both cases, if the randomly chosen neghboring
site is already occupied then the resident and incomer particles annihilate
each other leaving an empty site behind. 

In the case of $z=1$ the graph consists of disjoint pairs and the particles
vanish for $P>0$, while the particles survive on the single occupied pairs
if $P=0$. For $z=2$ the graph becomes a set of disjoint loops
and the feature of BARW can be described by the one-dimensional results
\cite{CT98,TT,J93}.
Our analyses will be concentrated on the random graphs with sufficiently
large $N$ and $z \ge 3$. Locally these graphs are similar to trees.
A distance between two sites can be introduced as the length (number of
steps) of the shortest path joining them. The average distance between
two sites increases logarithmically with the number of sites for
large $N$ \cite{Boll}.

The stationary state of this system is characterized by the average
concentration of walkers ($c$) that will be determined by using different
methods. For a locally tree like structure the generalization of the
one-dimensional dynamical cluster technique is straightforward \cite{gmf}.
In this case the particle distribution is described by the configuration
probabilities $p_k(n_1, \ldots ,n_k)$ ($n_i=0$ or 1) on the clusters
of the neighboring $k$ sites. Here we assume that these quantities satisfy
some symmetry (translation, rotation, reflection) and compatibility
relations. The one-point configuration probabilities are directly related
to the average concentrations, namely, $p_1(1)=c$ and $p_1(0)=1-c$.
Introducing an additional parameter $q$, the two-point configuration
probabilities are given as $p_2(1,1)=q$, $p_2(1,0)=p_2(0,1)=c-q$ and
$p_2(0,0)=1-2c+q$. Further parameters are required for $k>2$.

In the present case the time variation of $p_k$ can be expressed by the
terms of $p_k$ and $p_{k+1}$. For example,
\begin{equation}
\dot{p}_1(1)=(1-P) p_1(1) - p_2(1,1) + p_2(0,1)
\label{eq:mf1p}
\end{equation}
where we have summed the contribution of all the elementary processes
mentioned above. Notice that this equation is satisfied by the
absorbing state ($c=0$). At the level of one-point approximation
we assume that $p_2(n_1,n_2)=p_1(n_1) p_1(n_2)$. In this case the
non-trivial stationary solution of Eq. (\ref{eq:mf1p}) obeys a simple
form,
\begin{equation}
c^{(1p)} = {P \over 2}
\label{eq:sol1p}
\end{equation}
independent of $z$.
At the level of $k$-point approximation the corresponding set of
equations is solved by using the Bayesian relations ($p_{k+1}$s are
approximated by the product of $p_k$ terms) \cite{gmf}. 
In the two-point approximation the straightforward calculation
gives the following stationary solution:
\begin{equation}
c^{(2p)} = P { 2(z-2)-(z-3)P \over 4(z-1) - 2 z P +2 P^2 } \, .
\label{eq:sol2p}
\end{equation}
This result refers to the absence of pair correlations in the limit
$z \to \infty$ as well as at $P=1$ for any values of $z$. At higher
levels the stationary solutions are evaluated numerically.

In order to check these results we have performed MC simulations on
random graphs with $N=500,000$ sites varying the branching rate $P$ for
$z=4$ and 3. The simulations are started from a randomly half-filled
graph and the concentration is obtained by averaging over $10^4$ MC steps
per particles after some thermalization.

\begin{figure}
\centerline{\epsfxsize=8cm
            \epsfbox{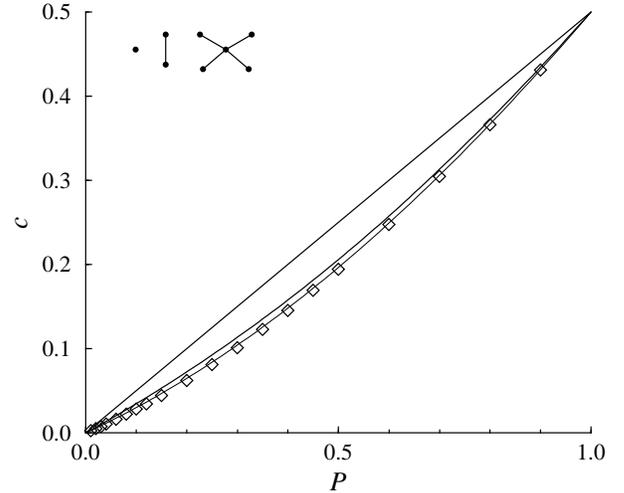}
            \vspace*{2mm}   }
\caption{The average concentration of particles as a function of
branching rate for $z=4$. The symbols represent MC data, the solid curves
comes from the 1-, 2- and 5-point approximations (from top to bottom)
on clusters indicated at the top.}
\label{fig:barwg4}
\end{figure}

Figure \ref{fig:barwg4} compares the MC data to the prediction of
$k$-point approximations for $z=4$. Here the results of 3- and 4-point
approximations are omitted because their deviation from $c^{(2p)}$ is
comparable to the line thickness. In this case $c=AP$ when
$P \to 0$. The $A$ coefficients obtained by MC simulation and
5-point approximation are slightly different, namely, $A^{(MC)}=0.250(2)$
and $A^{(5p)}=0.278(1)$.

\begin{figure}
\centerline{\epsfxsize=8cm
            \epsfbox{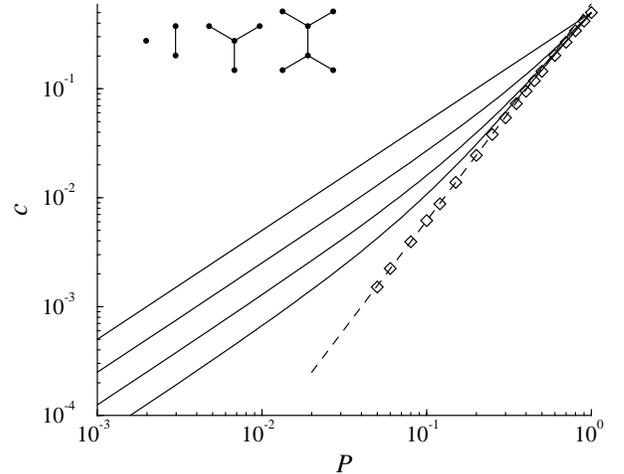}
            \vspace*{2mm}   }
\caption{Log-log plot of the average concentration of particles {\it vs} $P$
if $z=3$. The solid curves indicate the prediction of generalized
mean-field methods at the levels of 1-, 2-, 4-, and 6-point approximations
(from top to bottom). The fitted curve (dashed line) on the MC data
(open diamonds) shows quadratic behavior.}
\label{fig:barwg3}
\end{figure}

In the light of pair approximation [see Eq.~(\ref{eq:sol2p})] better and
better agreement is expected when increasing $z$. For $z=3$, however,
significant differences can be observed between the MC results and
the prediction of $k$-point approximations when $P \to 0$. To magnify the
discrepancy between the two methods the $P$-dependence of the
average concentration is illustrated in a log-log plot. As shown
in Fig.~\ref{fig:barwg3} the MC data refer to a quadratic behavior
for small $P$ values, meanwhile the $k$-point approximations predict
linear behavior with a coefficient decreasing when $k$ is increased.

Here it is worth mentioning that the MC data have remained unchanged
within the statistical error (comparable to the line thickness) when
the random graph was generated in a different way. This investigation
was motivated by the small word model suggested by Watts and
Strogatz \cite{WS}. In this case the points with the  first and
second joints form a single loop and the third ones are chosen randomly.

The trend in the prediction of $k$-point approximations implies the
possibility that the coefficient of the linear term goes to zero in the
limit $k \to \infty$. Now we introduce a modified mean-field theory
taking explicitly into consideration the mutual annihilation of the 
parent and his offspring as mentioned above. It will be shown that the
vanishment of the linear term is directly related to fact that half of the 
branching process yields mutual annihilation.

In this description the $R(z)$ probability of this mutual annihilation
process is approximated with the probability of returning to the
starting site for a single random walker on the same graph.
This simplification is exact in the limits $P$ and $c \to 0$
when the random walks of the parent and his offspring are practically not
affected by other particles. The appearance of additional offsprings during
the recurrence is considered a second order process whose effect
will be discussed below.

The main advance of this approach is that we can use the exact results
obtained for the simple random walk on the Bethe lattices which are
locally similar to the present random graphs.
Hughes and Sahimi \cite{HS} and  Cassi \cite{Cassi89}
have obtained that $R(z)=1/(z-1)$ and the average number of steps to return
to the starting site is $\tau(z)=2(z-1)/(z-2)$. The value of $\tau(z)$
indicates that the particle visits only a few sites before its recurrence
and during this short period it is not capable to distinguish the Bethe
lattices from the present random graphs due to the absence (low probability)
of loops. In fact, this is the reason why our analysis is restricted
to large $N$.

\begin{figure}
\centerline{\epsfxsize=8cm
            \epsfbox{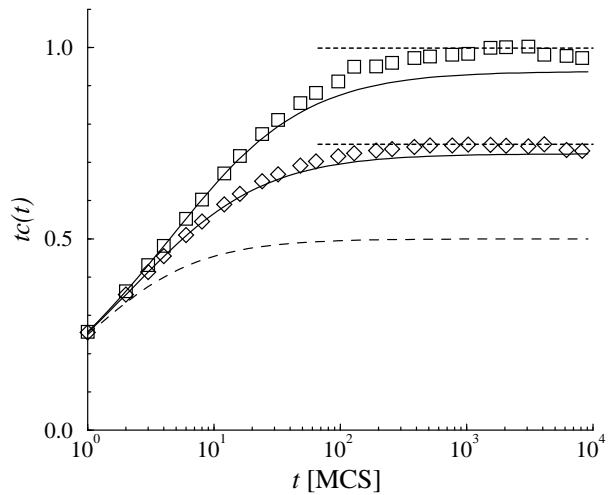}
            \vspace*{2mm}   }
\caption{Time-dependence of average concentration multiplied by time for $P=0$
and $c(0)=1/2$. The open squares (diamonds) represent MC data for $z=3$ ($z=4$).
The upper (lower) solid curve indicates the prediction of 6-point (5-point)
approximation for $z=3$ ($z=4$). The dashed line shows the prediction of classical
mean-field approximation. The fitted asymptotic functions are represented
by dotted lines.}
\label{fig:arwt}
\end{figure}

As a result of the mutual annihilation, $R(z)$ portion of the branching
process can be considered as spontaneous annihilations. On the other hand,
$1-R(z)$ portion of the branching events results in ``independent''
particles. These features can be easily built into a modified mean-field
theory which obeys the following form:

\begin{eqnarray}
\dot{p}_1(1)=&-&(1-P) p_1(1) - p_2(1,1) - P R(z) p_2(1,0) \nonumber \\
             &+&[1-P R(z)] p_2(0,1) \, .
\label{eq:mmf}
\end{eqnarray}
The two-point configuration probabilities are also affected by the
random walk itself, because a given particle leaves an empty site behind
when it steps to one of the neighboring sites. Consequently, the value of
$p_2(1,0)$ is larger than those predicted by the mean-field theory.
The above mentioned techniques confirm that this correlation can be well
described by a simple parameter defined as
\begin{equation}
p_2(1,1)=Q(z) p_1(1) p_1(1)
\label{eq:Q}
\end{equation}
where $Q(z) \le 1$ and the remaining two-point configuration probabilities
are determined by the compatibility conditions. The value of $Q(z)$ is
related to the asymptotic time-dependence for the annihilating random walk
($P=0$) when the particle concentration $c(t)$ decreases monotonously.
From Eqs.~(\ref{eq:mmf}) and (\ref{eq:Q}) we obtain that
\begin{equation}
c(t) = {1 \over 2 Q(z) t}
\label{eq:ct}
\end{equation}
when $t \to \infty$. Figure \ref{fig:arwt} compares this
result with the MC data as well as with the prediction of $k$-point
approximations obtained by numerical integration of the corresponding
set of differential equations for $z=3$ and 4. In the MC simulations
the system is started from a half-filled random initial state for $N=10^6$
and the data are averaged over 50 runs. The classical mean-field
(1-point) approximation corresponds to the choice of $Q(z)=1$. From the
$k$-point approximations, however, we can deduce smaller $Q(z)$ values. For
example, in the pair approximation $Q^{(2p)}(4)=3/4$ and $Q^{(2p)}(3)=2/3$.
These estimations can be improved if we choose larger and larger clusters as
illustrated in Fig.~\ref{fig:arwt}. We have obtained $Q^{(5p)}(4)=0.692(1)$
for $z=4$ and $Q^{(6p)}(3)=0.534(1)$ for $z=3$ using 5- and 6-point
approximations respectively. At the same time, the numerical fitting to
the MC data results in $Q^{(MC)}(4)=0.67(2)$ and $Q^{(MC)}(3)=0.50(2)$.
Notice that the predictions of $k$-point approximations tend slowly toward
the MC values. According to the numerical integration in $k$-point
approximations the asymptotic behavior is independent of the initial
concentration $c(0)$.

Using the expression (\ref{eq:Q}) the nontrivial stationary solution of
Eq.~(\ref{eq:mmf}) obeys the following simple form:
\begin{equation}
c = P {1 - 2 R(z) \over 2 [1 - P R(z)] Q(z)} \, .
\label{eq:mmfsol}
\end{equation}
Evidently, beside this expression Eq.~(\ref{eq:mmfsol}) has a trivial
solution ($c=0$) which remains the only one for $R > 1/2$.
In the limit $P \to 0$ the leading term of Eq.~(\ref{eq:mmfsol})
can be expressed as $c=P[1-2R(z)]/2Q(z)$. This prediction 
coincides with the above mentioned MC result for $z=4$
if $Q^{(MC)}(4)$ is substituted for $Q(z)$.

According to Eq.~(\ref{eq:mmfsol}) the average concentration
becomes zero for $z=3$. This is a consequence of the fact that the
rare branching processes do not modify the average number of walkers
because these events result in zero or two walkers with the same
probability on a longer time scale. There exists,
however, a second order term neglected above which is responsible for
the quadratic behavior. Namely, during the recurrence one of the
(parent and offspring) particles can create an additional walker with
a probability proportional to $P^2R(z)[\tau(z)-1]$ and this event
reduces the rate of spontaneous annihilation [$PR(z)$] in
Eqs.~(\ref{eq:mmf}) and (\ref{eq:mmfsol}). Unfortunately, this rough
approach can not reproduce accurately the coefficient of the quadratic term
because of the simplification we used in the derivation of
Eq.~(\ref{eq:mmf}).

The present analysis indicates that the probability of recurrence of a
single walker, the long time behavior of the time-dependent concentration
for the annihilating random walks
and the stationary concentration for the BARW in the limit $P \to 0$ are 
strongly related to each other. The modified mean-field theory
suggests a way to describe this relation on random regular graphs
for sufficiently large number of sites. The relation is based
on two simple conditions. Namely, $R(z) \le 1/2$ and $\tau(z)$ is finite.
These conditions are also fulfilled on the cubic lattices for dimensions
$d>4$ \cite{remark} where the rigorous analysis predicts mean-field
type behavior \cite{CT98}. Conversely, the modified mean-field theory is
not adequate for lower dimensions ($d \le 3$) because $\tau = \infty$.
It is expected, however, that the present analysis can be extended to
the investigation of BARW on such random graphs characterized with
different probability distributions of connectivity \cite{ER,BA}.
The present random graphs for $z=3$ represent a particular situation
where the coefficient of the linear term vanishes as a consequence of
$R(3)=1/2$. The solutions of the modified mean-field theory imply the 
the possibility of another interesting situation for those above mentioned
random graphs \cite{Boll,WS,ER,BA} where the average value of $R$ is larger
than 1/2. In this case the particles can survive if the branching rate
exceeds a treshold value as it happens on the one- and two-dimensional
lattices.

\acknowledgements
Support from the Hungarian National Research Fund (T-23552) is acknowledged.


\begin{references}

\bibitem{BG}M. Bramson and L. Gray, Z. Wahrsch. verw. Gebiete {\bf 68},
447 (1985).

\bibitem{MD}J. Marro and R. Dickman, {\it Nonequilibrium Phase
Transitions in Lattice Models} (Cambridge University Press,
Cambridge, 1998).

\bibitem{CT96}J. L. Cardy and U. C. T\"auber, Phys.\ Rev.\ Lett.\
{\bf 77}, 4780 (1996).

\bibitem{CT98}J. L. Cardy and U. C. T\"auber, J.\ Stat.\ Phys.\ 
{\bf 90}, 1 (1998).

\bibitem{dpconj}H. K. Janssen, Z. Phys.\ B {\bf 42}, 151 (1981);
P. Grassberger, Z. Phys.\ {\bf 47}, 365 (1982).

\bibitem{J93}I. Jensen, J.\ Phys.\ A {\bf 26}, 3921 (1993); Phys.\ Rev.\
E {\bf 50}, 3623 (1994).

\bibitem{IT}N. Inui and A. Yu. Tretyakov, Phys.\ Rev.\ Lett.\ {\bf 80},
5148 (1998).

\bibitem{TT}H. Takayasu and A. Yu.\ Tretyakov, Phys.\ Rev.\ Lett.\ 
{\bf 68}, 3060 (1992).

\bibitem{Cassi92}D. Cassi, Phys.\ Rev.\ Lett.\ {\bf 68}, 3631 (1992).

\bibitem{MW}F. Merkl and H. Wagner, J. Stat.\ Phys.\ {\bf 75}, 153 (1994).

\bibitem{BCV}R. Burioni, D. Cassi, and A. Vezzani, Phys.\ Rev.\ E {\bf 60},
1500 (1999).

\bibitem{Boll}B. Bollob\'as, {\it Random Graphs} (Academic Press, New York,
1985).

\bibitem{gmf}H. A. Gutowitz, J. D. Victor, and B. W. Knight, Physica
{\bf 28D}, 18 (1987).

\bibitem{WS}D. J. Watts and S. H. Strogatz, Nature {\bf 393}, 440 (1998).

\bibitem{HS}B. D. Hughes and M. Sahimi, J. Stat.\ Phys.\ {\bf 29}, 781 (1982).

\bibitem{Cassi89}D. Cassi, Europhys.\ Lett.\ {\bf 9}, 627 (1989).

\bibitem{remark}Some quantitative features of random walk on Bethe and
cubic lattices ($R$ and $\tau$) are compared in Ref.~\cite{HS}.

\bibitem{ER}P. Erd\H os, A. R\'enyi, Publ.\ Math.\ Inst.\ Hun.\ Acad.\ Sci.\
{\bf 5}, 17 (1960).

\bibitem{BA}A.-L. Barab\'asi and R. Albert, Science {\bf 286}, 509 (1999).

\end{references}
\end{document}